\documentclass[12pt,a4paper]{article}
\usepackage[cp1251]{inputenc}
\usepackage[english]{babel}
\usepackage{epsfig}
\usepackage{amssymb}
\usepackage{amsmath}

\newcommand{\bm}[1]{{\boldsymbol{\bf #1}}}
\renewcommand{\vec}[1]{{\bm{#1}}}
\newcommand{\fr}[2]{{\dfrac{#1}{#2}}}
\newcommand{\sfr}[2]{{{#1}/{#2}}}

\newcommand{\pdiff}[2]{{\fr{\partial{#1}}{\partial{#2}}}}

\newcommand{\spdiff}[2]{{\sfr{\partial{#1}}{\partial{#2}}}}

\title%
{%
Magnetic-Field Structure in the Accretion Disks of Semi-Detached Binary Systems%
\footnote{Submitted in Astronomy Reports, 2010, \textbf{54}, No. 9.}
}%
\author%
{%
 A.G. Zhilkin$^{1,2}$\thanks{E-mail: zhilkin@inasan.ru}, \ %
 D.V. Bisikalo$^{1}$\\ %
\textit{\small $^{1}$ Institute of Astronomy, Russian Academy of Sciences, Moscow, Russia}\\%
\textit{\small $^{2}$ Chelyabinsk State University, Chelyabinsk, Russia}\\%
}%
\date{}

\begin{document}

\maketitle%

\begin{abstract}

\noindent The results of three-dimensional MHD numerical simulations are used to investigate the characteristic properties of the magnetic-field structures in the accretion disks of semi-detached binary systems. It is assumed that the intrinsic magnetic field of the accretor star is dipolar. Turbulent diffusion of the magnetic field in the disk is taken into account. The SS Cyg system is considered as an example. The results of the numerical simulations show the intense generation of a predominantly toroidal magnetic field in the accretion disk. Magnetic zones with well defined structures for the toroidal magnetic field form in the disk, which are separated by current sheets in which there is magnetic reconnection and current dissipation. Possible observational manifestations of such structures are discussed. It is shown that the interaction of a spiral precessional wave with the accretor’s magnetosphere could lead to quasi-periodic oscillations of the accretion rate. 
\end{abstract}

\section{Introduction}

There are a large number of close binary systems in which the magnetic field plays an important role in mass transfer and accretion. First and foremost, such systems include magnetic and intermediate polars, as well as X-ray binaries \cite{Warner1995, Campbell1997, Lipunov1987}. These are semi-detached binary systems in which one of the components (the donor) fills its Roche lobe, allowing material from its envelope to overflow to the other component (the accretor) through the inner Lagrange point $L_1$. The accretors in magnetic and intermediate polars are white dwarfs with surface magnetic fields of $10^7$-$10^8$~G and $10^4$-$10^6$~G, respectively. The accretors in X-ray binaries are neutron stars with intrinsic magnetic fields of $10^{12}$-$10^{13}$~G. It is believed that mass transfer in intermediate polars and X-ray binaries can lead to the formation of accretion disks around the compact objects \cite{Warner1995, Campbell1997}. The intrinsic magnetic field can appreciably influence the structure of the accretion disk and determine the character of accretion onto the star \cite{Koldoba2002, Romanova2003, Norton2004, Norton2008}. 

In our previous studies \cite{Zhilkin2009, ZhilkinASR2010}, we presented results of numerical simulations of the structure of MHD flows in semi-detached binary systems. These showed that the main parameters of the accretion disk, such as the accretion rate and characteristic density, can change when the magnetic field is taken into account. However, the structure of the magnetic field in the disk was not considered. The magnetic field in the accretion disk around a compact object can be strengthened due to differential rotation, radial motions and dynamo action. Diffusion, turbulent dissipation, and magnetic buoyancy can weaken the magnetic field. The action of these effects can lead to the formation of a magnetic field with a complex structure, since different effects may dominate in different parts of the disk. Note that, in contrast to individual objects (a star, accretion disk, etc.), an important role can be played in close binaries by specific mechanisms for the generation of magnetic field \cite{Dolginov1987}. For example, the Herzenberg dynamo may operate in magnetic polars  \cite{Herzenberg1958, Brandenburg1998}, leading to the generation of magnetic field in the envelope of the secondary. It is believed that the decelerating electromagnetic torque that arises in this case leads to synchronization of the rotation of the components in these systems \cite{Campbell1997}.

The generation of toroidal magnetic field due to differential rotation dominates in the inner part of the disk \cite{Campbell1997}. The character of the generated field is determined by the rotation law in the disk. However, effects associated with the presence of a poloidal velocity in the disk, which redistribute the magnetic field throughout the disk, may be important here \cite{BisnovatyKogan1974, BisnovatyKogan1976, Rothstein2008}. The magnetic field can interact with waves arising in the inner regions of the disk \cite{Bisikalo2004, Bisikalo2005}, leading to quasi-periodic variations in the accretion rate onto the star \cite{ZhilkinASR2010}.

In the outer part of the accretion disk, magnetic field may partially be generated by the dynamo mechanism. Both a laminar dynamo (due to non-axially symmetric motions) \cite{Braginsky1964}, and a turbulent $\alpha\omega$ dynamo \cite{Parker1982, Moffatt1978, Vainshtein1980} may operate in accretion disks in close binary systems. The dynamo generation of magnetic field in accretion disks has been considered, for example, in \cite{Parker1971, Pudritz1981, Stepinski1990, Rudiger1995, Campbell1997, Campbell2005}. The dynamo generation of magnetic field requires that the mean helicity $\alpha$ for non-axially symmetric and turbulent motions of the gas not display mirror symmetry about the equatorial plane of the disk. Physically, this means that the numbers of right-handed and left-handed vortices are not equal, since the Coriolis force creates additional winding of the vortices. When $\alpha>0$ the quadrupolar mode of the magnetic field is dominant above the equatorial plane \cite{Ruzmaikin1988}. However, detailed numerical simulations of magnetic turbulence \cite{Brandenburg1995, Ziegler2000} due to the development of magnetorotational instability in accretion disks \cite{Velihov1959, Balbus1998} show that the opposite situation can also be realized under certain conditions \cite{Brandenburg2002, Rekowski2003}. In this case, the dipolar component of the magnetic field becomes dominant. Note that this component of the field is preferred for the generation of bipolar outflows from the disk via the magnetocentrifugal mechanism \cite{Blandford1982}.

Here, we use three-dimensional numerical simulations to study the structures of magnetic fields in accretion disks in semi-detached binary systems. We consider the SS Cyg system as an example, whose white dwarf has a dipolar magnetic field. The numerical models take into account radiative heating and cooling, as well as diffusion of the magnetic field due to current dissipation in turbulent vortices and magnetic buoyancy.

This paper is organized as follows. Section 2 briefly describes the model and numerical method used. Section 3 presents the results of the numerical simulations. In Section 4, we consider a simple analytical model for magnetic-field generation in accretion disks in semi-detached binary systems. Our main results are summarized in the Conclusion.

\section{The model}

We will describe the flows in a semi-detached binary using a Cartesian coordinate system ($x$, $y$, $z$) in a frame rotating with angular velocity $\vec{\Omega}$ about the common center of mass. The coordinate origin is specified to be the center of the accretor, and the center of the donor is located a distance $A$ from the accretor on the $x$ axis. The $z$ axis is directed along the rotational axis of the system. 

The MHD equations can be written
\begin{equation}\label{eq2.2}
  \pdiff{\rho}{t} +
  \nabla \cdot \left( \rho\vec{v} \right) = 0,
\end{equation}
\begin{equation}\label{eq2.3}
  \pdiff{\vec{v}}{t} +
  \left( \vec{v} \cdot \nabla \right) \vec{v} =
  -\fr{1}{\rho} \nabla P -
  \fr{1}{4\pi\rho}
  \left( \vec{B} \times \left(\nabla \times \vec{B}\right)\right) +
  2\left(\vec{v} \times \vec{\Omega}\right) -
  \nabla\Phi,
\end{equation}
\begin{equation}\label{eq2.4}
  \pdiff{\vec{B}}{t} =
  \nabla \times
  \left( \vec{v} \times \vec{B} - \eta (\nabla \times \vec{B}) \right),
\end{equation}
\begin{equation}\label{eq2.5}
  \rho T
  \left(
  \pdiff{s}{t} + \left( \vec{v} \cdot \nabla \right) s
  \right) =
  n^2 \left(\Gamma - \Lambda \right) +
  \fr{\eta}{4\pi} \left(\nabla \times \vec{B}\right)^2.
\end{equation}
Here, $\rho$ is the density, $\vec{v}$ the velocity, $P$ the pressure, $\vec{B}$ the magnetic field, $s$ the entropy per unit mass of gas, $n = \rho/m_{\text{p}}$ the number density, $m_{\text{p}}$ the proton mass, $\eta$ the magnetic viscosity, and $\Phi$ the Roche potential. The radiative heating and cooling $\Gamma$ and $\Lambda$ have complex dependences on the temperature $T$ \cite{Cox1971, Dalgarno1972, Raymond1976, Spitzer1981}). Our numerical model uses a linear approximation for these functions in the vicinity of the equilibrium temperature $T = 11230~K$ \cite{Bisikalo2003, Zhilkin2009, ZhilkinMM2010}, corresponding to an effective temperature for the accretor of $37000~K$. The term $2\left(\vec{v} \times \vec{\Omega}\right)$ in the equation of motion (\ref{eq2.3}) describes the Coriolis force. The last term on the right-hand side of the entropy equation (\ref{eq2.5}) describes heating of the matter due to current dissipation. The density, entropy, and pressure are related by the ideal-gas equation of state, $s = c_V \ln(P / \rho^{\gamma})$, where $c_{V}$ is the specific heat of the gas at constant volume and $\gamma = 5/3$ is the adiabatic index.

We consider here a semi-detached binary system with a forming accretion disk and with synchronous rotation of the components, when the rotational period of the accretor is equal to the orbital period. We will also assume that the magnetic field of the accretor is dipolar, so that the field strength is determined by
\begin{equation}\label{eq2.6}
  \vec{B}_{*} =
  \fr{3\left(\vec{m}\cdot\vec{r}\right)\vec{r}}{r^5} -
  \fr{\vec{m}}{r^3},
\end{equation}
where $\vec{m}$ is the magnetic moment of the accretor. The direction of the magnetic moment $\vec{m}$ need not coincide with the direction of the angular velocity of the binary system $\vec{\Omega}$. The components of $\vec{m}$ are $m_x = m \sin\theta \cos\phi$, $m_y = m \sin\theta \sin\phi$, and $m_z = m \cos\theta$, where $m$ is the magnitude of $\vec{m}$, $\theta$ the angle between $\vec{m}$ and the $z$ axis, and $\phi$ the angle between the $x$ axis and the projection of $\vec{m}$ onto the $xy$ plane.

Our numerical modeling takes into account the effects of magnetic-field diffusion \cite{Zhilkin2009, ZhilkinASR2010}. Turbulent diffusion of the magnetic fields in accretion disks is determined by two main effects \cite{Campbell1997}. The first is magnetic reconnection and dissipation of currents in turbulent vortices. The second is associated with buoyancy of flux tubes of the toroidal magnetic field generated in the disk as a result of differential rotation. Both these effects are included in our model. Note that the magnetic viscosity coefficient due to magnetic buoyancy depends on the strength of the generated magnetic field. Therefore, the turbulent diffusion of magnetic field in the disk as a whole is non-linear.

We used the Nurgush three-dimensional parallel numerical code for the numerical simulations \cite{ZhilkinMM2010}. This code is based on a higher-order, Gudonov-type, finite-difference scheme. The original unified-variable technique for the MHD equations \cite{Zhilkin2007} makes it possible to use an adaptive grid in the numerical code. Our computations used a geometrically adaptive grid that became denser in the equatorial plane and toward the surface of the accretor. This enabled us to appreciably increase the resolution of the vertical structure of the accretion disk and in the region of the accretor magnetosphere. In order to minimize the numerical errors in the difference scheme, we computed only the magnetic field induced by currents in the accretion disk and in the outer envelope \cite{Tanaka1994, Powell1999}. We used the eight-wave method \cite{Powell1999, Dellar2001} to clean the divergence of the magnetic field. 

The equations describing the diffusion of the magnetic field are non-linear. Therefore, the application of explicit methods to solve for this field would lead to too rigid constraints on the time step used. In our approach, this equation is solved numerically using an implicit, locally one-dimensional method with a factorized operator \cite{Samarsky1971}. Another difficulty is that this equation contains mixed derivatives in the curvilinear, non-orthogonal coordinate system due to the geometrically adaptive grid used. Our method solves this problem via regularization of the factorized operator. The factorization procedure essentially reduces to replacing the co-multiplicative operators comprising the initial factorization operator with some equivalent tridiagonal operators. The regularization parameter is determined by the maximum modulus of the eigenvalue of the metric tensor describing the curvilinear coordinate system. The non-linear terms are correctly taken into account in the scheme via an iterative process that is applied until a solution with a specified accuracy is obtained. At each iteration, a system of linear, algebraic equations with a tridiagonal matrix arises, which is solved numerically using a scalar fitting method.

\section{Computational results}

We have investigated the structure of the magnetic field in the accretion disk using the results of numerical simulations of MHD flows in the SS Cyg system \cite{Zhilkin2009, ZhilkinASR2010}. The donor in this system is a red dwarf with mass $0.56~M_{\odot}$, while the accretor is a white dwarf with mass $0.97~M_{\odot}$. The orbital period of the system is $P_{\text{orb}} = 6.6$~hr, and its semi-major axis is $A=2.05R_{\odot}$ \cite{Giovannelli1983}. In our computations, the magnetic field at the surface of the white dwarf was specified to $10^5$~G \cite{Fabbiano1981, Kjurkchieva1999}, with the orientation of the magnetic axis given by $\theta=30^{\circ}$ and $\phi=0^{\circ}$. The speed of gas flowing at the inner Lagrange point $L_1$ was taken to be equal to the local sound speed, $c_s = 7.4~\text{km}/\text{s}$, and the corresponding effective temperature of the donor to be $4000~K$. The gas density at $L_1$ was taken to be $\rho(L1) = 1.1 \times 10^{-7}~\text{g}/\text{cm}^{3}$, and the mass-transfer rate to be $\dot{M} = 10^{-9}~M_{\odot}/\text{year}$. The solution was found in the region ($-0.56A \le x \le 0.56A$, $-0.56A \le y \le 0.56A$, $-0.28A \le z \le 0.28A$) on the geometrically adaptive grid \cite{ZhilkinMM2010}. 

\begin{figure}[t]
\centering
\includegraphics[width=8cm]{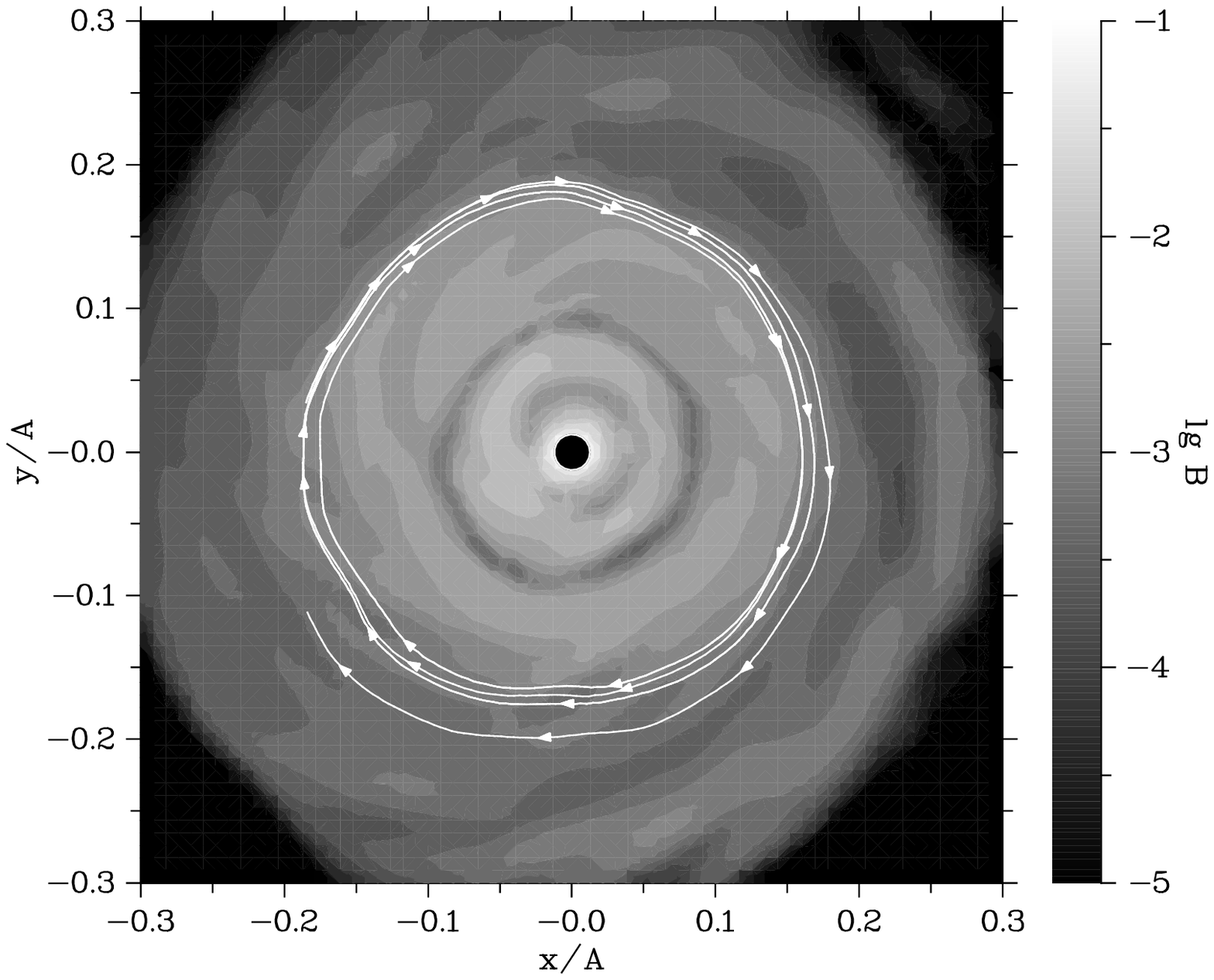}%
\hspace{0.05cm}
\includegraphics[width=8cm]{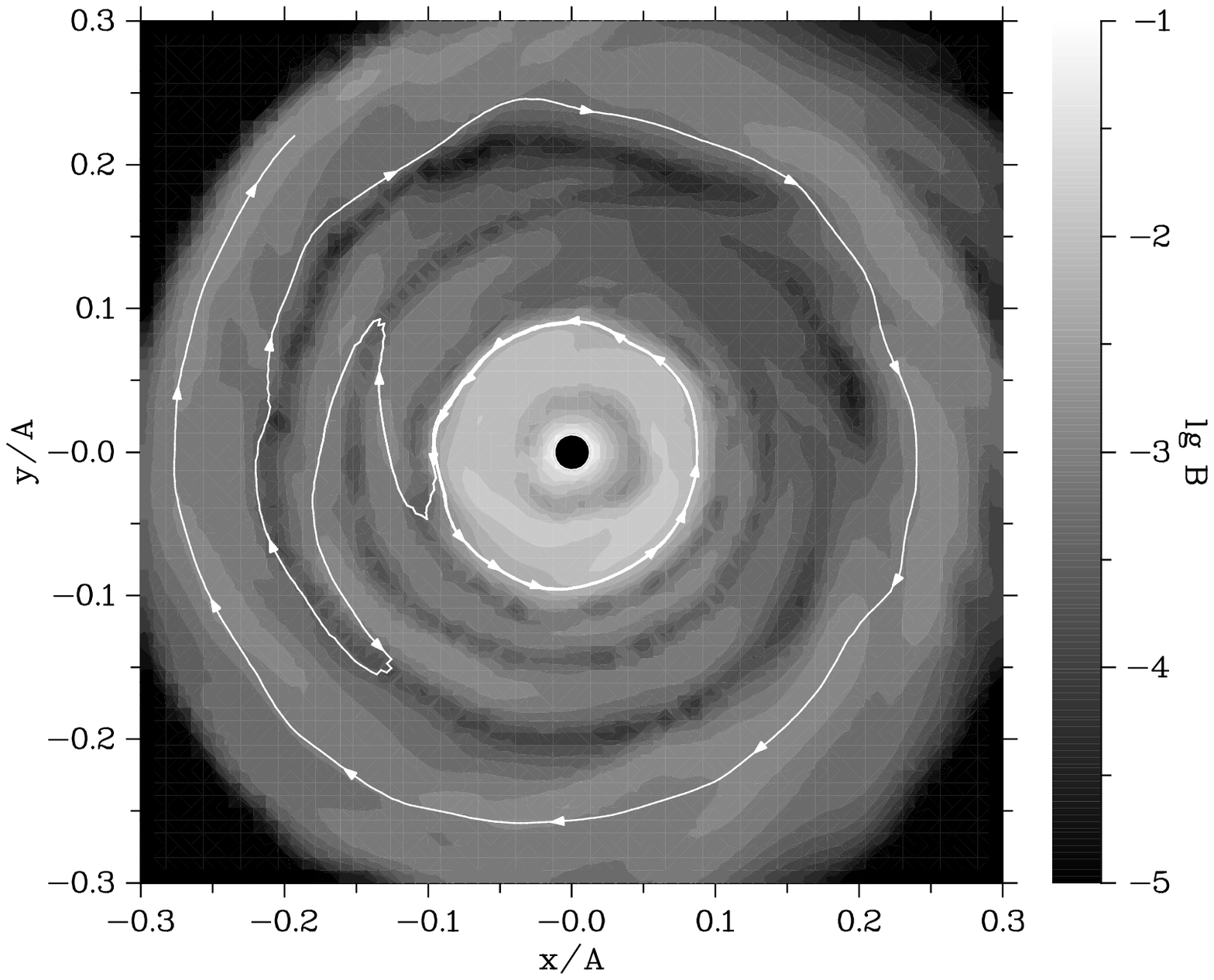}%
\caption{Magnetic-field distribution in the equatorial plane of the disk at times $12.75 P_{\text{orb}}$ (left) and $13.36 P_{\text{orb}}$ (right). The light line with the arrows shows the magnetic field passing through the point $x=-0.175A$, $y=0.00A$.}%
\label{fig_lgb}
\end{figure}

\begin{figure}[t]
\centering
\includegraphics[width=8cm]{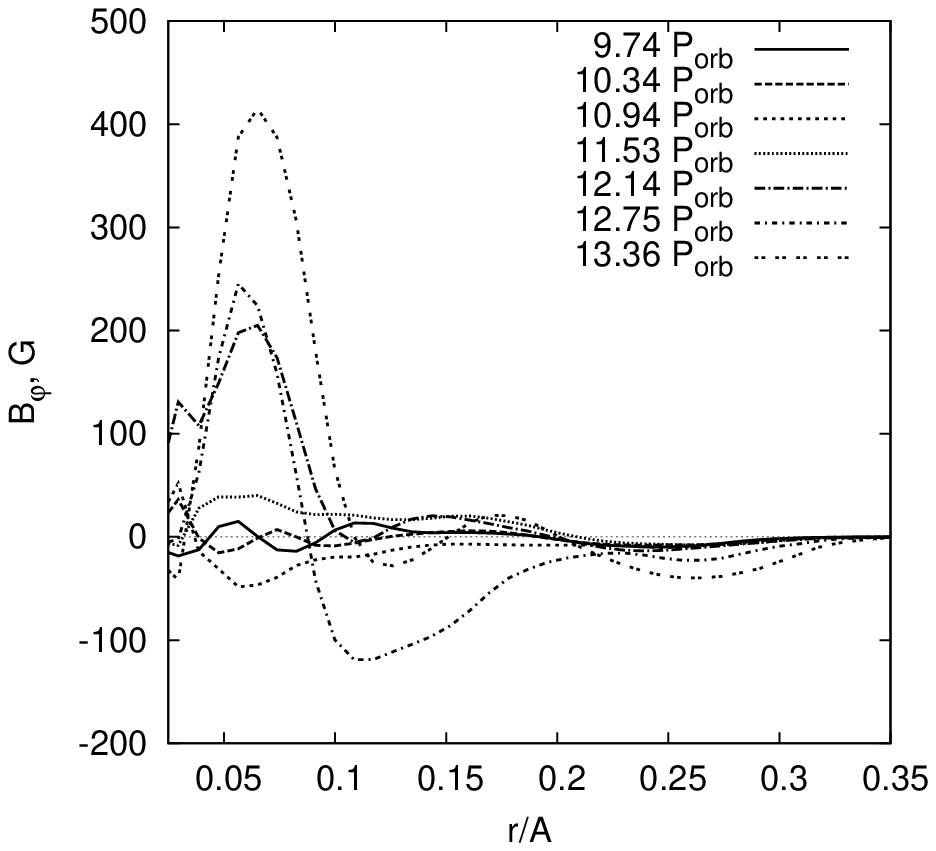}%
\hspace{0.05cm}
\includegraphics[width=8cm]{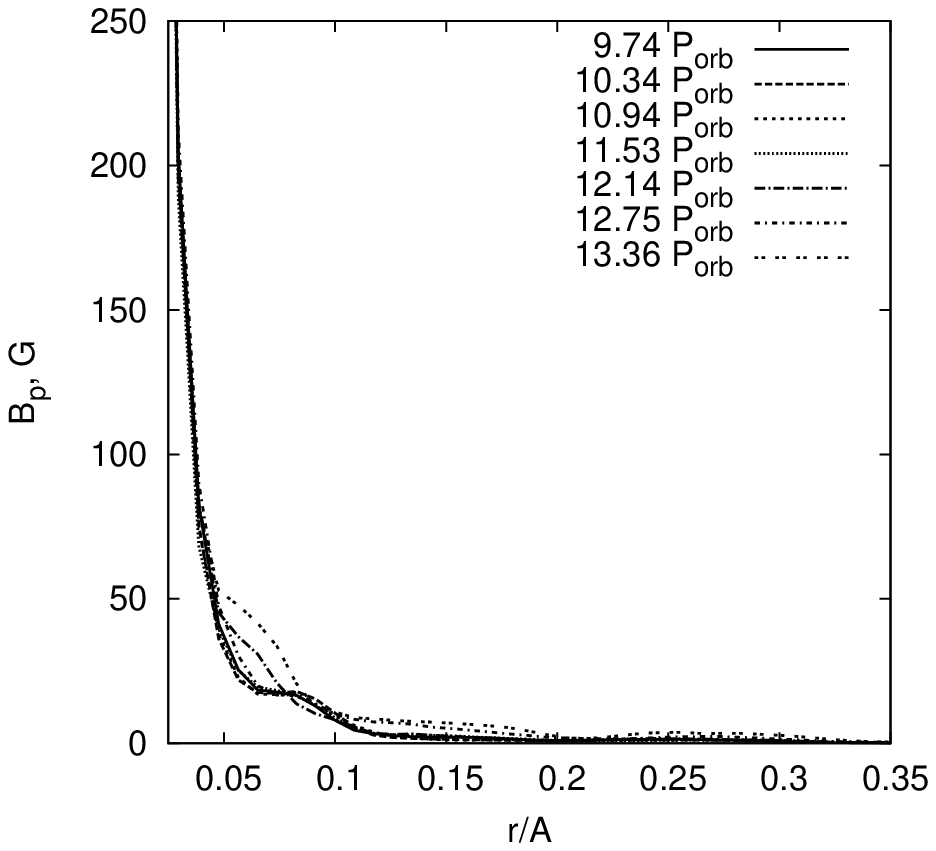}%
\caption{Radial distribution of the azimuth-averaged toroidal $B_{\varphi}$ (left) and poloidal $B_p$ (right) magnetic fields in the disk for various times.}%
\label{fig_bn}
\end{figure}

The structure of the magnetic field in the accretion disk obtained in the numerical simulations is shown in Figs. \ref{fig_lgb} and \ref{fig_bn}. Figure \ref{fig_lgb} shows the distribution of the magnetic field $B$ (in units of $\sqrt{4\pi\rho(L_1)}A\Omega$) in the equatorial ($xy$) plane of the disk for times of $12.75 P_{\text{orb}}$ (left) and $13.36 P_{\text{orb}}$ (right). The light curve with arrows indicates the magnetic-field line passing through the point $x=-0.175A$, $y=0.00A$. The left-hand panel in Fig. \ref{fig_bn} shows the radial distribution of the azimuth-averaged toroidal field in the disk $B_{\varphi}$ for the various times, while the right-hand panel in Fig. \ref{fig_bn} shows the analogous distribution of the poloidal field $B_p = \sqrt{B_r^2 + B_z^2}$.

Inspection of these figures indicates that three zones can be clearly distinguished in the disk: an inner zone where there is an intense generation of toroidal field due to the differential rotation of the disk, a zone of current sheets, and an outer zone of dissipation of the magnetic field. The radius of the inner zone is roughly $0.1A$. The current sheets are visible in Fig. \ref{fig_lgb} as dark rings. The magnetic-field lines change direction on either side of a current sheet. No fewer than three current sheets can be distinguished in the right-hand panel of Fig. \ref{fig_lgb} The current-sheet zone is located at distances from approximately $0.1A$ to $0.2A$. In Fig. \ref{fig_bn}, the current sheets correspond to points where the toroidal field $B_{\varphi}$ vanishes. There is no intense generation of field by differential rotation in the outer zone, at distances exceeding $0.2A$, since the stellar magnetic field is very weak in this region. However, the right-hand panel of Fig. \ref{fig_bn} shows that a weak poloidal field is generated in this zone. Overall, the magnetic field in the accretion disk is predominantly toroidal.

The formation of current sheets occurs at the boundary of, or even inside, the inner zone. The change in the sign of  $B_{\varphi}$ is associated with a variation in the rotation law in the disk. A transition zone where there is an intense loss of angular momentum forms near the star. In this zone, the angular velocity of the gas decreases from the Keplerian velocity to the rotational velocity of the stellar magnetic-field lines. This variation of the rotation law influences the generation of the toroidal field.

\begin{figure}[t]
\centering
\includegraphics[width=4cm]{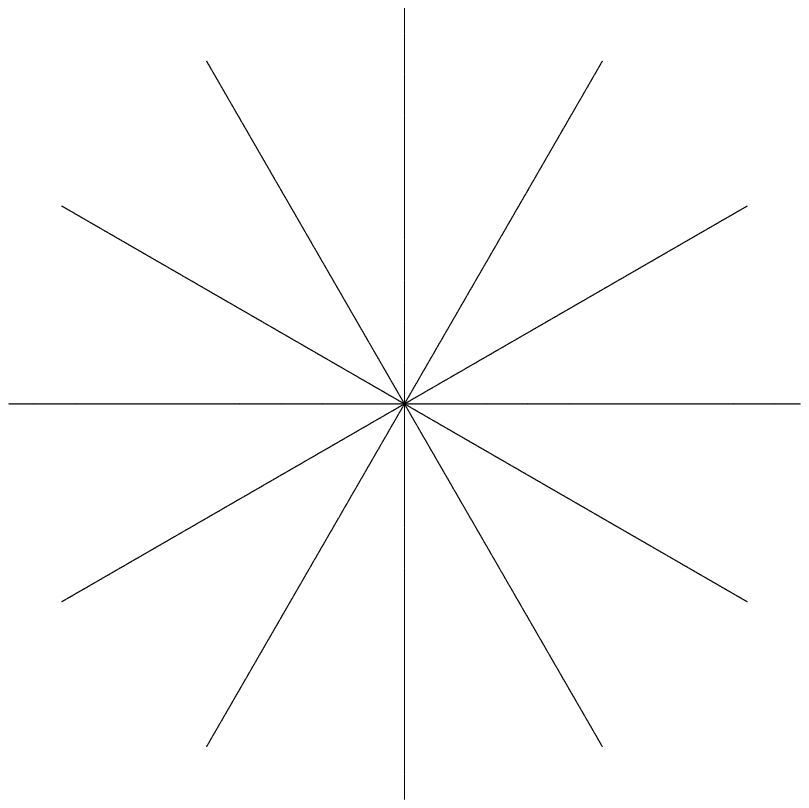}%
\hspace{0.05cm}
\includegraphics[width=4cm]{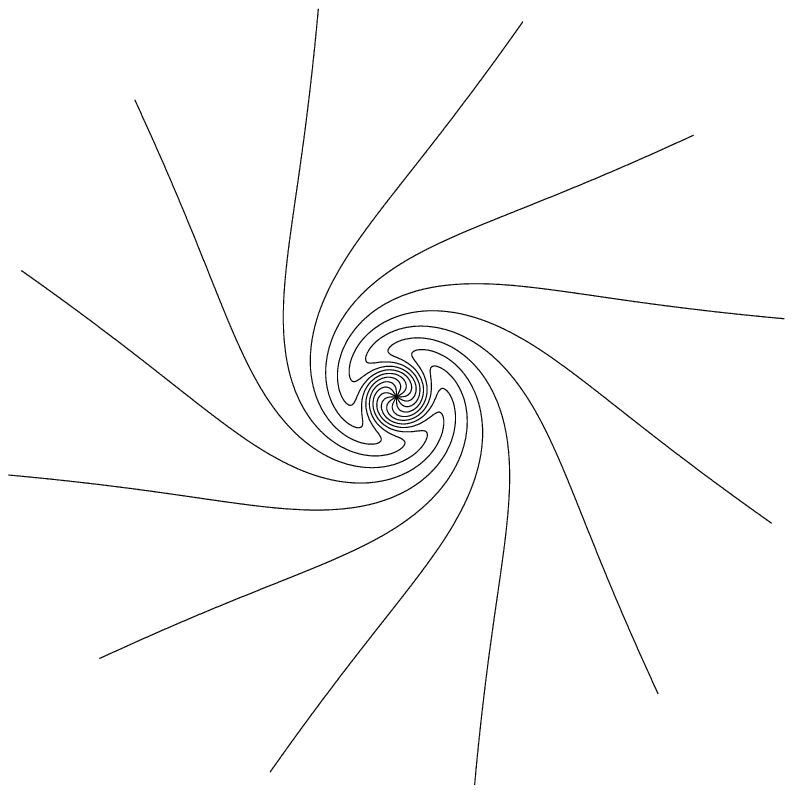}%
\hspace{0.05cm}
\includegraphics[width=4cm]{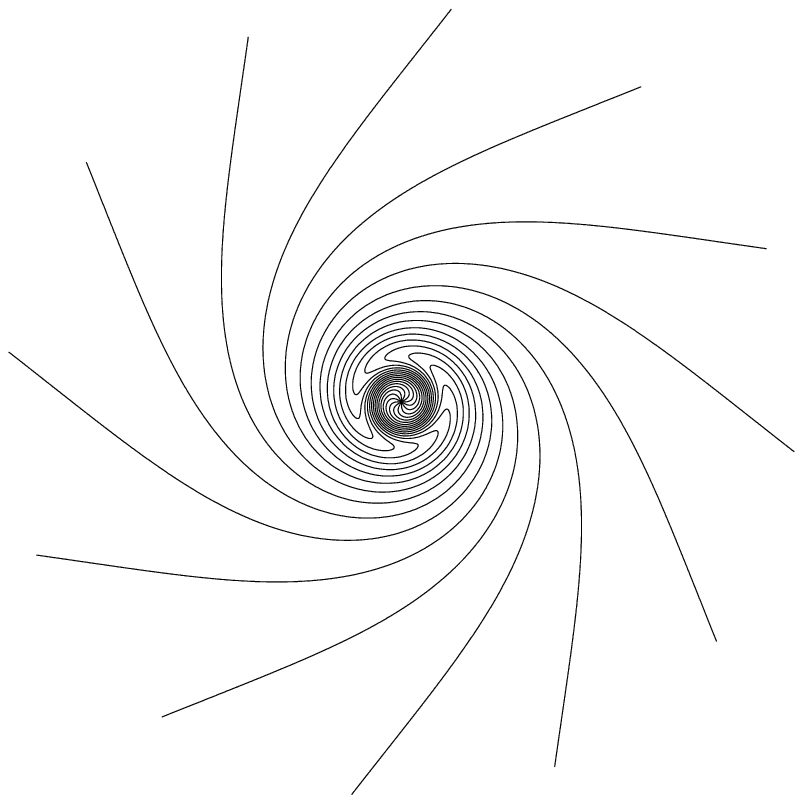}%
\hspace{0.05cm}
\includegraphics[width=4cm]{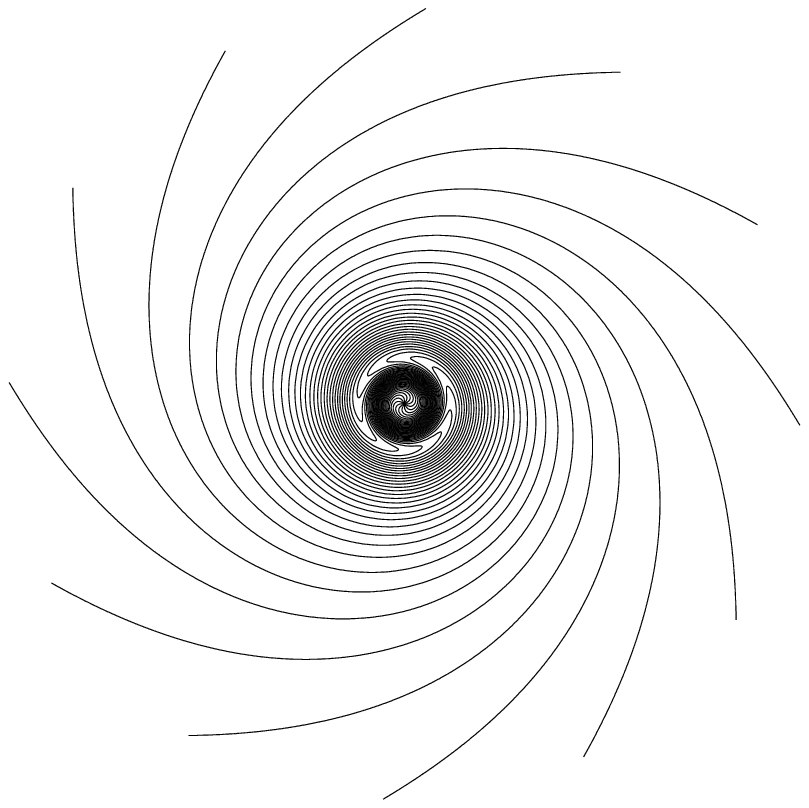}%
\caption{Visual model for the formation of two magnetic zones and current sheets in an accretion disk. Schematics of the magnetic-field lines in the equatorial plane are shown for (left to right) the initial time and after one, two, and five revolutions of the disk.}%
\label{fig_mf}
\end{figure}

These arguments can be clarified using the following picture. Consider the model dependence for the angular velocity $\omega$ on radius $r$:
\begin{equation}\label{eq3.1}
	\omega = 
	\fr{\omega_0 (\sfr{r}{r_0})^a}{1 + (\sfr{r}{r_0})^{a + 3/2}},
\end{equation}
where $r_0$ is some characteristic radius, approximately equal to the radius of the magnetosphere. We obtain when $r \ll r_0$ the asymptotic dependence $\omega = \omega_0 (\sfr{r}{r_0})^a$, which determines the rotation law in the region of the magnetosphere. In the opposite limit when $r \gg r_0$, expression (\ref{eq3.1}) leads to a Keplerian rotation law, $\omega = \omega_0 (\sfr{r}{r_0})^{-3/2}$. Figure \ref{fig_mf} illustrates the evolution of the magnetic field and formation of a current sheet. Figure \ref{fig_mf} (first) shows the projection of the stellar magnetic-field lines onto the equatorial plane at the initial time. The subsequent panels show the pattern of the magnetic-field lines in the disk obtained by transforming the initial field pattern using the rotation law (\ref{eq3.1}) at times $\sfr{2\pi}{\omega_0}$ (second), $\sfr{4\pi}{\omega_0}$ (third) and $\sfr{10\pi}{\omega_0}$ (fourth), respectively. Figure \ref{fig_mf} shows that the current sheet and magnetic zones can already clearly form after the first several revolutions of the disk. However, the subsequent evolution of these structures can appreciably influence the dissipation of the magnetic field, radial motions, and dynamo processes.

It follows that no current sheets should form in the case of equilibrium rotation of the star \cite{Lipunov1987}, when the corotation radius (distance where the rotational velocities of the field lines and of the disk material are equal) is equal to the magnetosphere radius. However, the disk material in such a star should be strongly wound up. Current sheets should form in systems where the corotation radius exceeds the magnetosphere radius. This condition is associated with the relationship between the accretor’s rotational angular velocity, magnetic field, and accretion rate. Our computations assumed that the accretor rotated synchronously, so that this condition is obviously satisfied. Note that the generation of large-scale magnetic fields in galaxies by the dynamo mechanism together with the presence of a complex rotation law can also lead to the formation of magnetic zones separated by current sheets \cite{Ruzmaikin1988}. Such a current ring is present in the galaxy M31, for example, at a radius of approximately 3 kpc.

\begin{figure}[t]
\centering
\includegraphics[width=8cm]{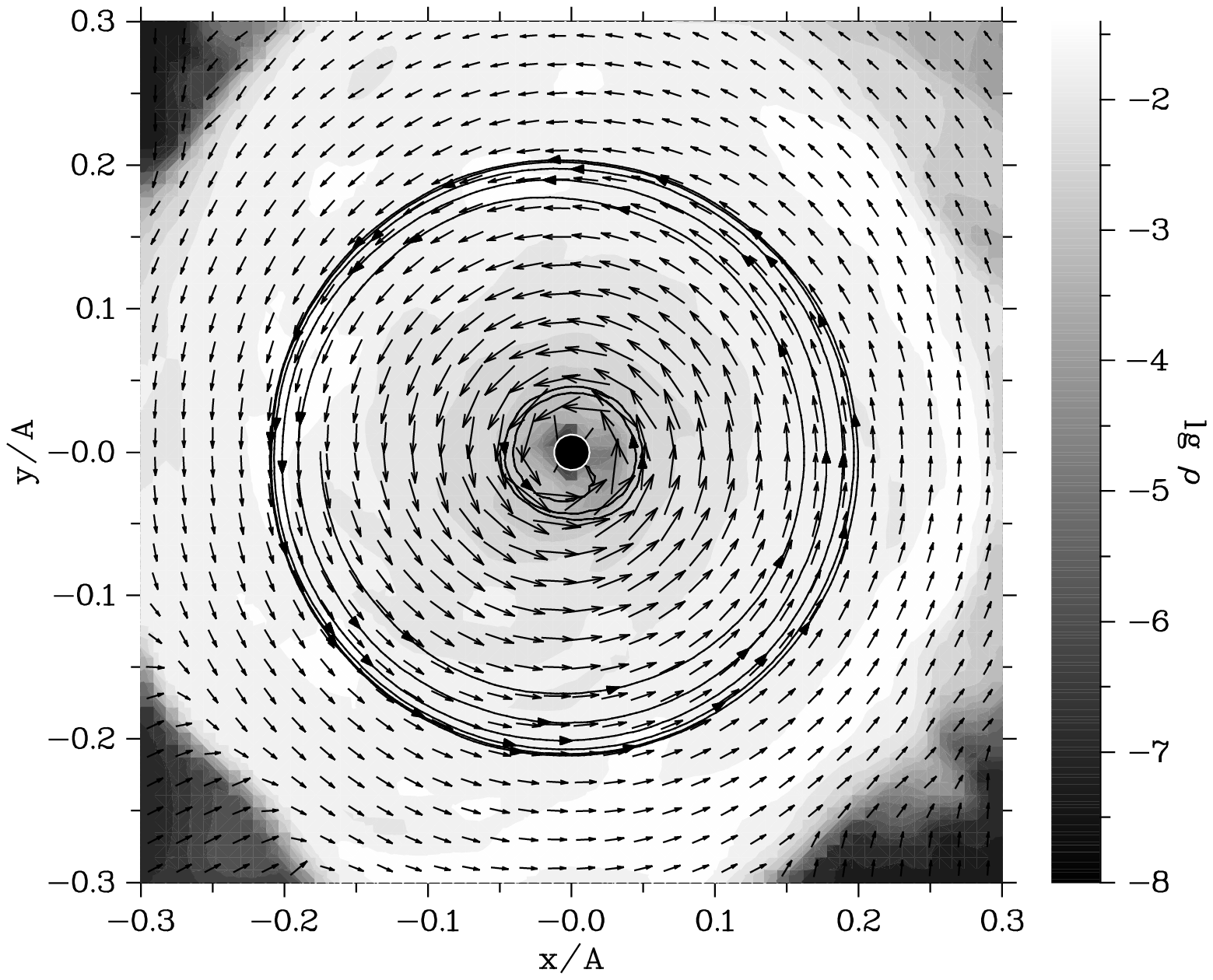}%
\hspace{0.05cm}
\includegraphics[width=8cm]{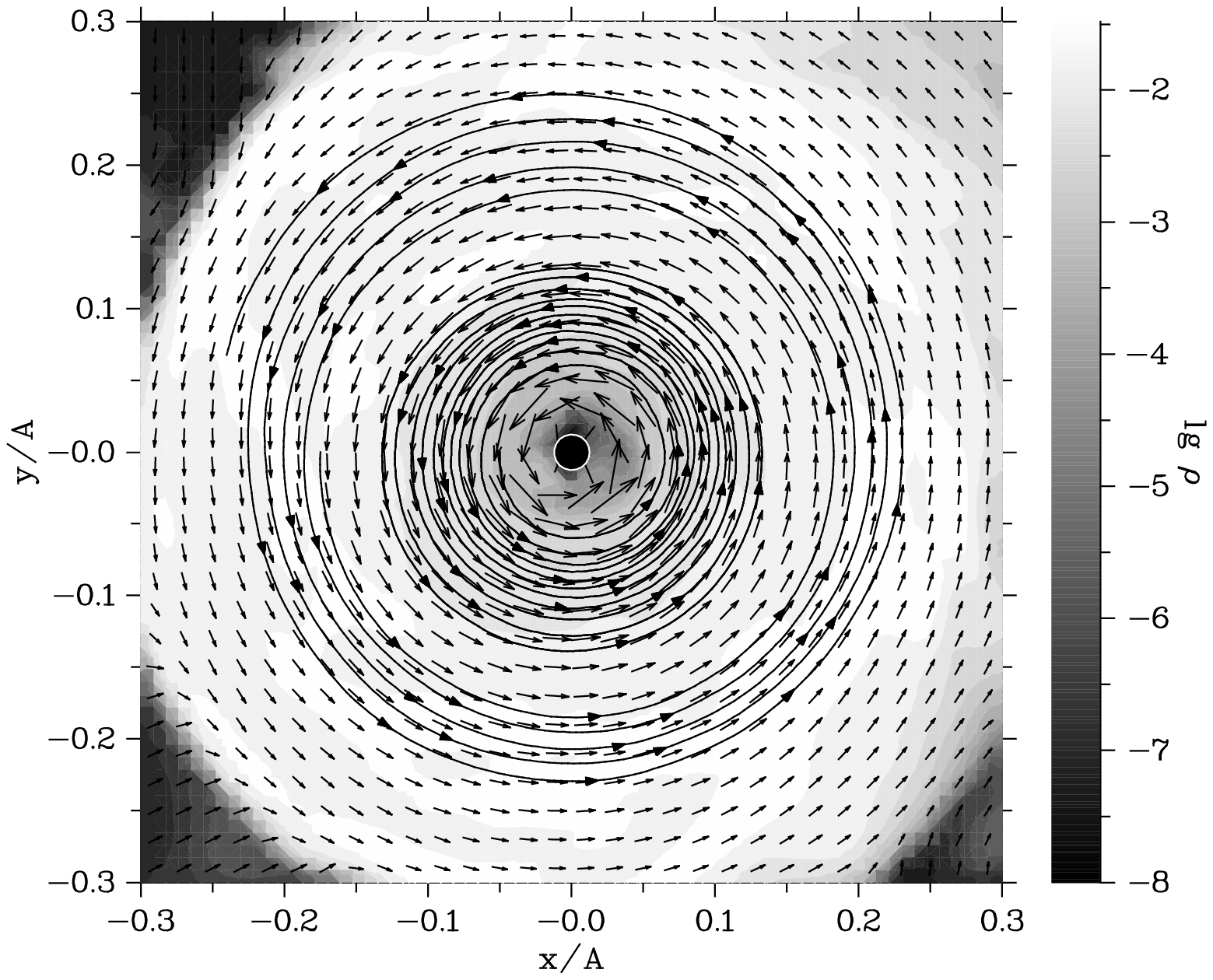}%
\caption{Distribution of the density (gray scale) and velocity (arrows) in the equatorial plane of the disk at times $12.75 P_{\text{orb}}$ (left) and $13.36 P_{\text{orb}}$ (right). The lines with arrows show stream lines emerging from the points  $x=-0.175A$, $y=0.00A$ and $x=-0.05A$, $y=0.00A$.}%
\label{fig_lgrho}
\end{figure}

\begin{figure}[t]
\centering
\includegraphics[width=8cm]{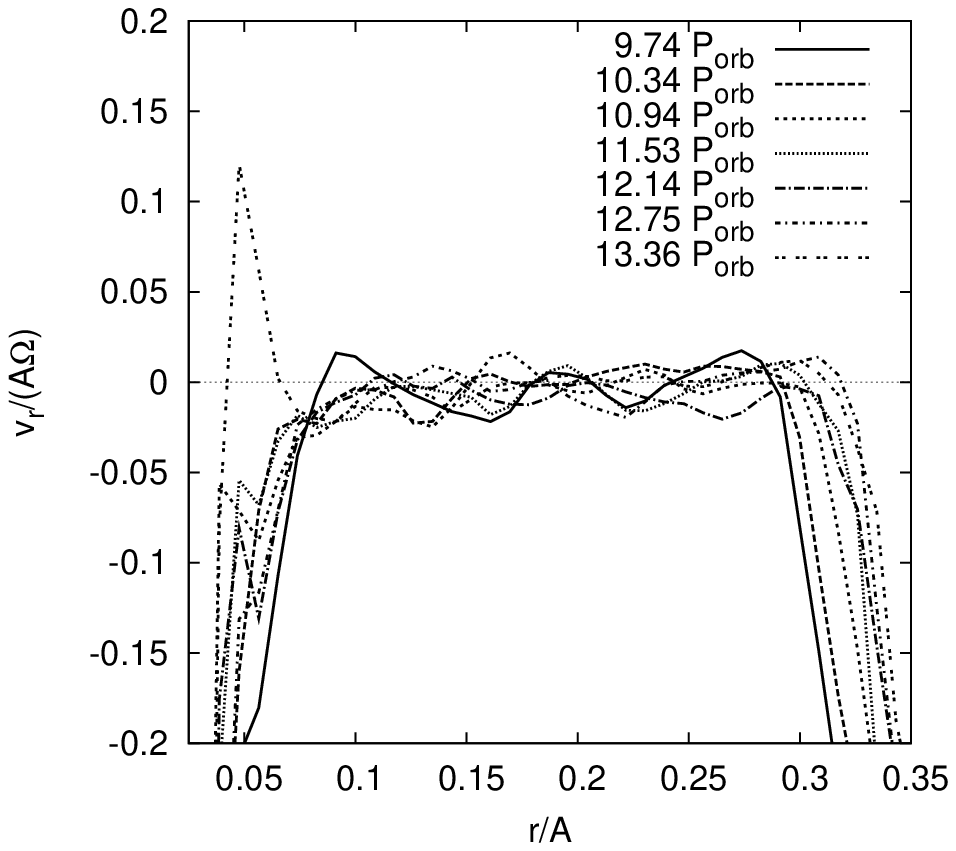}%
\caption{Radial distribution of the azimuth-averaged radial velocity in the disk $v_r$ for various times.}%
\label{fig_vr}
\end{figure}

The formed current sheet is carried to the outer part of the disk via decretion of the disk material. After some time, a new current sheet is formed in its place. Thus, several current sheets can exist simultaneously in the disk. Figure \ref{fig_lgrho} shows the distribution of the density (gray scale) and velocity (arrows) in the equatorial plane of the disk at times $12.75 P_{\text{orb}}$ (left) and $13.36 P_{\text{orb}}$ (right). The lines with arrows show stream lines emerging from the points $x=-0.05A$, $y=0.00A$ (inner zone) and $x=-0.175A$, $y=0.00A$ (outer zone). A comparison of the figure panels indicates that an alternation of accretion and decretion regimes occurs in the disk (especially its inner parts). The distribution of the azimuth-averaged radial velocity in the disk at various times is presented in Fig. \ref{fig_vr}, which clearly shows zones of accretion and decretion and their evolution with time.

The alternation of accretion and decretion regimes in the inner parts of the disk can be understood if the magnetic field is quasi-periodic. In fact, the radial gradient in the pressure due to the toroidal magnetic field is increased by the generation of this field, ultimately causing the accretion to cease. This should lead to a flow of the field to outer parts of the disk, decreasing the magnetic pressure, and giving rise to a transition back to an accretion regime. The variations in the field distribution in the inner zone can be seen in Figs. \ref{fig_lgb}, \ref{fig_bn}. 

\begin{figure}[t]
\centering
\includegraphics[width=8cm]{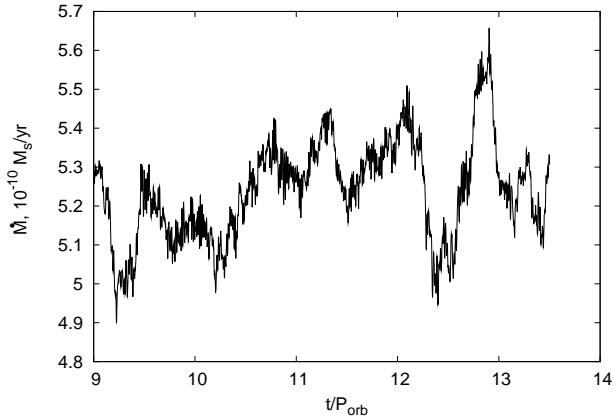}%
\caption{Evolution of the accretion rate onto the white dwarf.}%
\label{fig_mdot}
\end{figure}

The generated magnetic field interacts intensely with spiral waves in the inner zone. Of most interest is the interaction of the magnetic field with a spiral precessional wave that can arise in the inner, unperturbed parts of the disk due to retrograde precession of the binary orbit \cite{Bisikalo2004}. In the gas-dynamical case, the increase in the radial flow of matter behind this wave leads to a growth in the accretion rate and the formation of a compact region of enhanced energy release at the surface of the accretor. In the magnetic case, the accretion acquires a column-like character. Consequently, the accretion rate should grow at times when the precessional wave approaches the stellar surface near the magnetic poles. In the coordinate system considered, the rotational period of the spiral wave is approximately equal to the orbital period of the system, if we disregard the longer precessional period. Thus, approximately twice per period, we should observe flares in the accretion rate, during the passage of the spiral wave by the North and South magnetic poles of the star. Figure \ref{fig_mdot} shows the evolution of the accretion rate onto the star over several orbital periods. The behavior shown is in good agreement with this picture. The maximum amplitude of fluctuations in the accretion rate is about $15\%$. Note that this amplitude can reach $200\%$ in the case of ideal magnetic gas dynamics without diffusion effects \cite{ZhilkinASR2010}. The observed flux curves of SS Cyg in the UV between outbursts display similar behavior \cite{Giovanelli1999}, with the flux fluctuations reaching $70\%$.

\section{Generation of magnetic field}

\subsection{Basic equations}

To elucidate the main regularities in the generation of magnetic field in the accretion disk, we consider a simpler model in which the magnetic-dipole axis of the star coincides with its rotational axis. In this case, we can assume axial symmetry when describing the magnetic field averaged over the turbulent pulsations and the azimuth \cite{Zhilkin2009}. The averaged magnetic field\footnote{We will denote the averaged magnetic field in this section $\vec{B}$. This should not lead to confusion, since the initial, unaveraged field that satisfies \eqref{eq2.4}, is not used anywhere in this section.} can conveniently be represented as $\vec{B} = \vec{B}_{*} + \vec{b}$, where $\vec{B}_{*}$ is the magnetic field of the star and  $\vec{b}$ the magnetic field generated by inductional currents in the accretion disk. The initial induction equation  (\ref{eq2.4}) taking into account the dynamo effect can be written
\begin{equation}\label{eq4.1}
	\pdiff{\vec{b}}{t} = 
	\nabla \times
	\left(
	 \vec{v} \times \vec{B}_* +
	 \vec{v} \times \vec{b} +
	 \alpha \vec{b} -
	 \eta (\nabla \times \vec{b})
	\right).
\end{equation}
Since the magnetic field is poloidal, ($\vec{B}_{*\varphi} = 0$, $\vec{B}_{*p} = \vec{B}_{*}$), the toroidal component of the generated field is $\vec{b}_{\varphi} = \vec{B}_{\varphi}$. We denote $\vec{b}_p = \nabla \times (A\vec{n}_{\varphi})$, where $\vec{n}_{\varphi}$ is a unit vector in the azimuthal direction. Using this notation, Eq. (\ref{eq4.1}) can be rewritten as the system
\begin{equation}\label{eq4.2}
	\pdiff{A}{t} +
	\fr{1}{r}(\vec{v}_p \cdot \nabla)(rA)	= 
	\alpha B_{\varphi} + 
	\eta \left(\nabla^2 A - \fr{A}{r^2} \right),
\end{equation}
\begin{equation}\label{eq4.3}
	\pdiff{B_{\varphi}}{t} = 
	r(\vec{B}_p \cdot \nabla)\omega - 
	r \nabla \cdot \left(\fr{B_{\varphi}}{r}\vec{v}_p\right) -
	\fr{1}{r}\nabla\alpha \cdot \nabla(rA) -
	\alpha \left(\nabla^2 A - \fr{A}{r^2} \right) +
	\eta \left(\nabla^2 B_{\varphi} - \fr{B_{\varphi}}{r^2} \right),
\end{equation}
where $\omega = v_{\varphi}/r$ is the rotational angular velocity in the disk.

Under the conditions in the accretion disk, these equations can be appreciably simplified. Since $|v_r|, |v_z| \ll |v_{\varphi}|$ in the accretion disk, we can set $\vec{v}_p = 0$. Moreover, we can omit the negligibly small third and fourth terms in the right-hand side of (\ref{eq4.3}), which correspond to the $\alpha^2$ dynamo. The radial derivatives can be neglected in the diffusion terms compared to the vertical derivatives, since their ratio is $z/r \ll 1$. Further, using the estimates $\spdiff{\omega}{z} \approx \sfr{z}{r}\spdiff{\omega}{r}$, $|b_z\spdiff{\omega}{z}| \ll |b_r\spdiff{\omega}{r}|$ and
\begin{equation}\label{eq4.4}
  B_{*r} = \fr{3}{2} B_a  \left(\fr{R_a}{r}\right)^3 \fr{z}{r}, \ \ 
  B_{*z} = -\fr{1}{2} B_a  \left(\fr{R_a}{r}\right)^3, 
\end{equation}
we find
\begin{equation}\label{eq4.5}
  r(\vec{B}_p \cdot \nabla)\omega = \fr{z}{r} g B_0 - g \pdiff{A}{z},
\end{equation}
where $B_a$ is the magnetic field at the stellar surface, $g = r\spdiff{\omega}{r}$ a measure of the differential rotation,
and $B_0 = B_a  (\sfr{R_a}{r})^3$ the characteristic stellar magnetic field in the disk. Finally, we will assume that $g$ and $\eta$ depend only on $r$. 

These simplifications lead to the system
\begin{equation}\label{eq4.6}
	\pdiff{A}{t} = \alpha B_{\varphi} + \eta \pdiff{^2A}{z^2},
\end{equation}
\begin{equation}\label{eq4.7}
	\pdiff{B_{\varphi}}{t} = 
	\fr{z}{r} g B_0 - g \pdiff{A}{z} +
	\eta \pdiff{^2B_{\varphi}}{z^2}.
\end{equation}
These equations have a simple physical meaning. Equation (\ref{eq4.6}) describes the evolution of the poloidal magnetic field in the disk. The first term on the right-hand side determines the generation of poloidal field from toroidal field via the dynamo effect, while the second term determines the dissipation of the poloidal magnetic field. Equation (\ref{eq4.7}) describes the evolution of the toroidal magnetic field in the disk. The first and second terms on the right-hand side of this equation determine the generation of toroidal field from the stellar magnetic field and the generated poloidal field due to the differential rotation of the disk. The last term on the right-hand side describes the dissipation of the toroidal magnetic field.

\subsection{Generation of field by differential rotation}

The dynamo effect is weak in the inner parts of the disk, and magnetic field is generated by differential rotation in the disk. In this case, we can neglect the $\alpha B_{\varphi}$ term in (\ref{eq4.6}). As a result, we find $A = 0$, so that only toroidal magnetic field will be generated. The remaining Eq. (\ref{eq4.7}) acquires the form
\begin{equation}\label{eq4.8}
	\pdiff{B_{\varphi}}{t} = \fr{z}{r} g B_0 + 
	\eta \pdiff{^2B_{\varphi}}{z^2}.
\end{equation}

The general solution of (\ref{eq4.8}), satisfying the initial condition $B_{\varphi}(t = 0) = 0$ and the boundary conditions $B_{\varphi}(z = 0) = B_{\varphi}(z = H) = 0$ can be written \cite{Tikhonov1966}
\begin{equation}\label{eq4.9}
	B_{\varphi} = 
	2 B_{\varphi,0}
	\sum\limits_{n = 1}^{\infty}
	\fr{(-1)^n}{(\pi n)^3}
	\left[ 1 - e^{-(\pi n)^2 \sfr{t}{t_d}} \right]
	\sin\left( \fr{\pi n z}{H} \right),
\end{equation}
where $B_{\varphi,0} = \sfr{g B_0 H t_d}{r}$, and $t_d = \sfr{H^2}{\eta}$ is the diffusion time. In early stages of the field generation when $t \ll t_d$, the toroidal field grows linearly, $B_{\varphi} = \sfr{g B_0 z t}{r}$. This solution also describes the generation of field in the ideal MHD case with $\eta = 0$. Note, however, that this leads to a discontinuity in $B_{\varphi}$ at the surface of the disk at $z = H$. This means that, near this surface, an important role is played by dissipation of the field, so that this cannot be neglected, at least in this region. In the other limiting case when $t \gg t_d$ we obtain the stationary solution
\begin{equation}\label{eq4.10}
	B_{\varphi} = 
	\fr{1}{6} B_{\varphi,0}
	\fr{z}{H} \left( 1 - \fr{z^2}{H^2} \right).
\end{equation}
The maximum magnetic field $B_{\varphi}$ is reached at the point $z = \sfr{H}{\sqrt{3}}$. The mean field averaged over height
in the disk is $\bar{B}_{\varphi} = \sfr{B_{\varphi}}{24}$. 

The coefficient of $B_{\varphi,0}$ in (\ref{eq4.10}) describes the radial structure of the generated toroidal field. Let us
consider the model dependence (\ref{eq3.1}) of the rotational angular velocity in the disk. If we assume that $H \propto r$ and $\eta \propto r$, we can obtain the asymptotic relations $B_{\varphi,0} \propto r^{a - 2}$ as $r \ll r_0$ and $B_{\varphi,0} \propto r^{-7/2}$ as $r \gg r_0$. The behavior $B_{\varphi,0}$ as $r \to 0$ depends appreciably on the parameter $a$, which characterizes the rotation of the disk in the region of the stellar magnetosphere: $B_{\varphi,0} \to \infty$ as $a < 2$, $B_{\varphi,0} \to \text{const}$ as $a = 2$, and $B_{\varphi,0} \to 0$ as $a > 2$.

\begin{figure}[t]
\centering
\includegraphics[width=8cm]{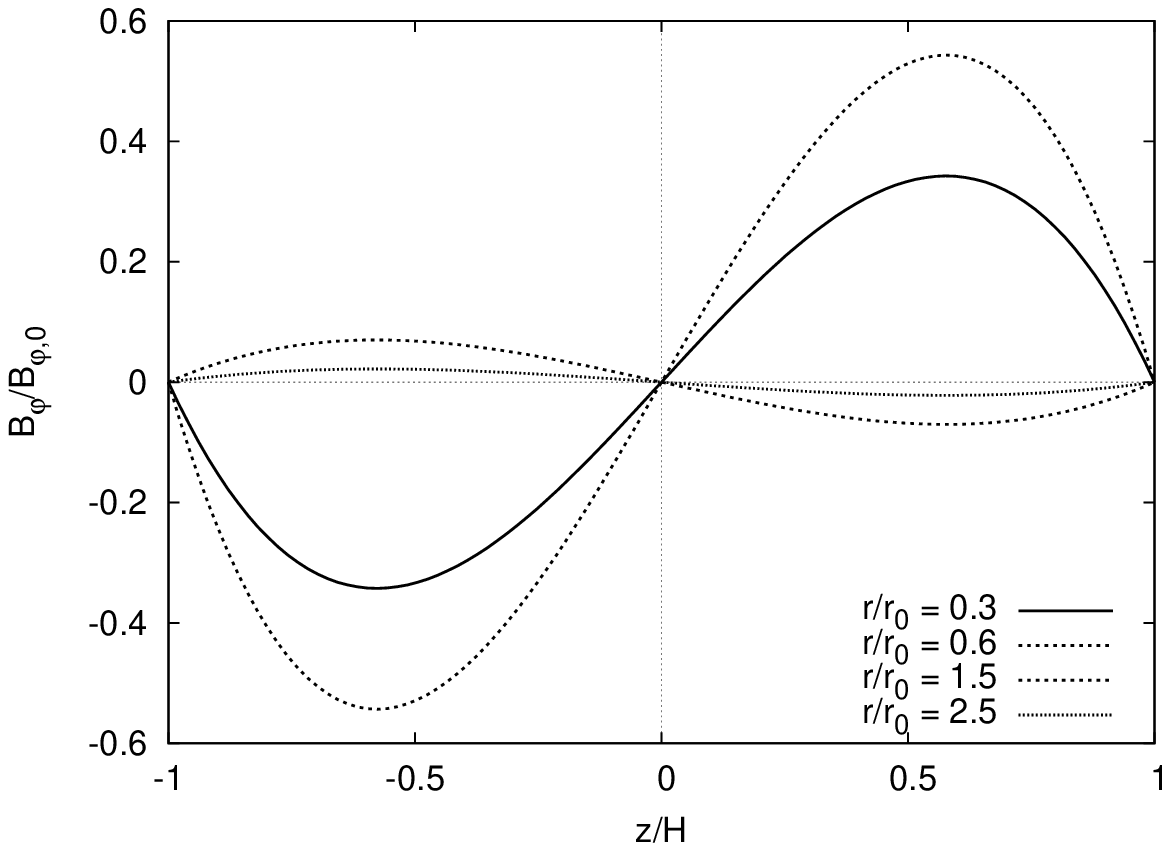}%
\hspace{0.05cm}
\includegraphics[width=8cm]{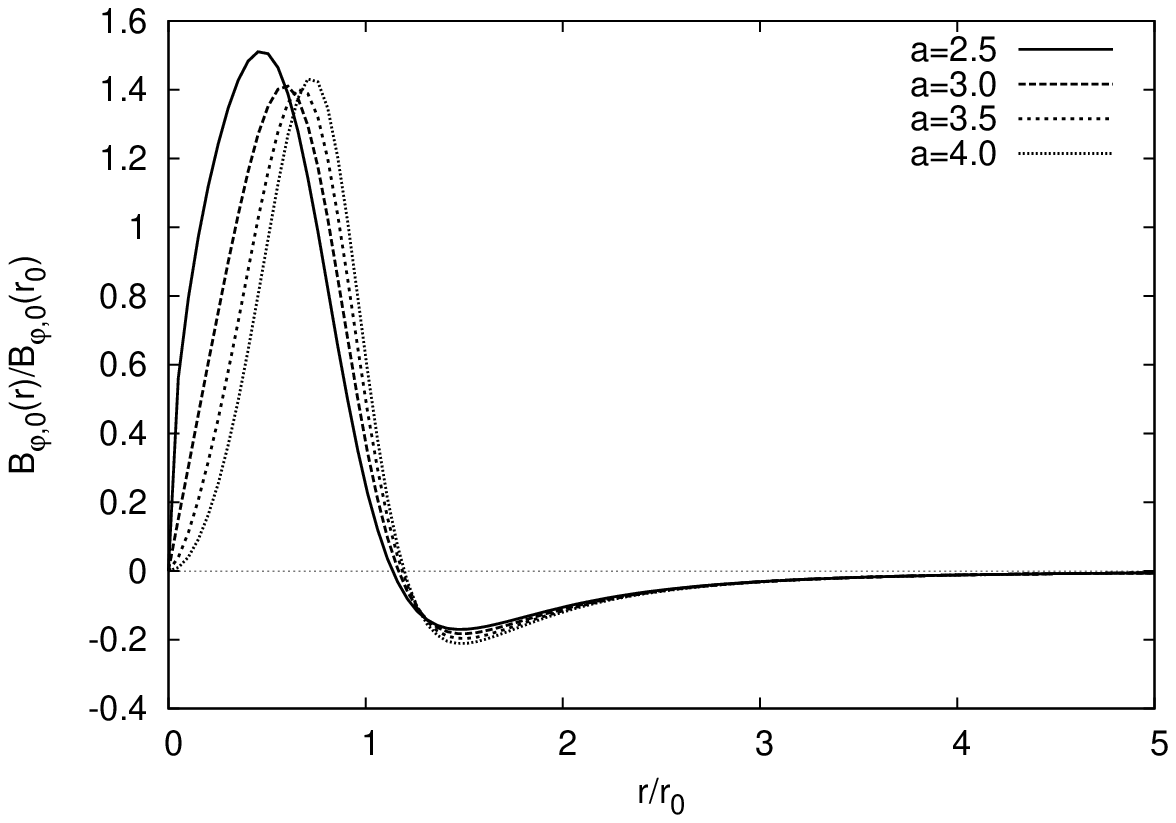}%
\caption{Vertical and radial profiles of the toroidal magnetic field in the inner parts of the accretion disk.}%
\label{fig_bphi}
\end{figure}

The left-hand panel in Fig. \ref{fig_bphi} shows characteristic vertical profiles of the stationary toroidal magnetic field in the accretion disk for $a = 3$ at various distances from the star. The right-hand panel of Fig. \ref{fig_bphi} shows the characteristic radial profiles of the field for various values of $a$. The sign of the magnetic field changes at the point $r = (\sfr{2a}{3})^{1/(a + 3/2)} r_0$. This means that neighboring magnetic-field lines will have opposite directions at this distance, leading to the formation of a current sheet separating two magnetic rings. The magnetic fields in the inner and outer magnetic rings have opposite signs. Recall that the numerical simulations lead to a similar picture (Fig. \ref{fig_bn}).

\begin{figure}[t]
\centering
\includegraphics[width=\textwidth]{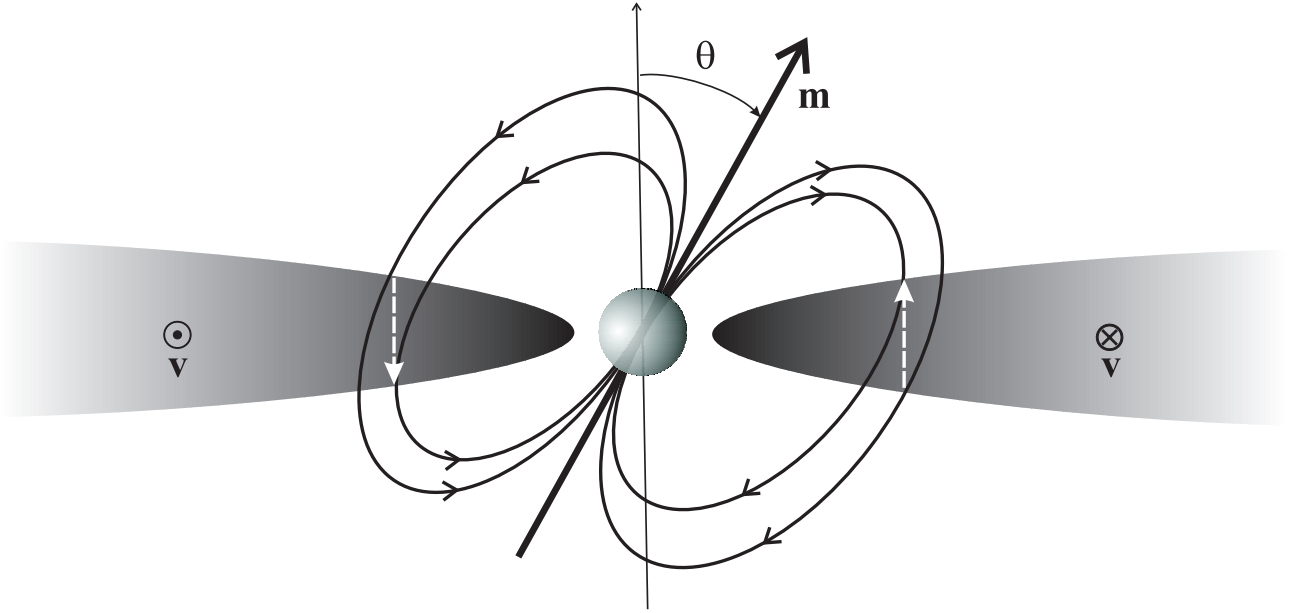}%
\caption{Illustration of the mechanism for the generation of a quadrupolar component of the toroidal field in the accretion disk in the case of an inclinedmagnetic moment.}%
\label{fig_mfgen}
\end{figure}

The antisymmetric character of the dependence of $B_{\varphi}$ on $z$ is related to the assumption that the magnetic and rotational axes of the star coincide. The antisymmetry of the field about the equatorial plane is disrupted if the magnetic axis is inclined. This situation is illustrated in Fig. \ref{fig_mfgen}: a clear vertical gradient of the field (shown by the light arrows) arises in the disk when the magnetic axis is inclined. In the right-hand panel, the field increases from top to bottom, and in the left-hand panel from bottom to top. As a result, primarily the stronger field will be amplified. Taking into account the rotation direction in the disk, the field directed toward the observer will grow more rapidly in the right-hand panel, and the field directed away from the observer in the left-hand panel. In both cases, the resulting toroidal field is positive. This type of toroidal-field distribution is also obtained in the computations. This effect will be expressed most strongly when the inclination of the magnetic axis is $\theta=45^{\circ}$. This reasoning bring us to the following conclusion. An antisymmetric toroidal field (dipolar mode) will be generated only when $\theta = 0^{\circ}$ and $90^{\circ}$. Any small deviation from these values will cause the quadrupolar mode to suppress the dipolar mode. Therefore, the generation of a purely dipolar mode in the accretion disks of intermediate polars may be unstable. The quadrupolar mode is more stable. Recall also that the dynamo generation of magnetic field in the outer part of the disk also leads primarily to the growth of the quadrupolar field component \cite{Ruzmaikin1988}.

\subsection{$\alpha\omega$ dynamo}

The dynamo effect can play an important role in the generation of magnetic field in the outer parts of the disk. Therefore, we must use the full system of equations (\ref{eq4.6}, \ref{eq4.7}) to describe the magnetic field. The quantity $\alpha$, in these equations is determined by the mean helicity of the velocity fluctuations $\delta\vec{v}$:
\begin{equation}\label{eq4.11}
	\alpha = -\fr{\tau}{3}
	\left\langle 
	\delta\vec{v} \cdot (\nabla \times \delta\vec{v})
	\right\rangle,
\end{equation}
where $\tau$ is the characteristic correlation time. The angular brackets denote an average over the ensemble of turbulent pulsations and over azimuth. In both averages, the characteristic correlation time can be estimated as $\tau = \sfr{H}{(\alpha_T c_s)}$. 

\begin{figure}[t]
\centering
\includegraphics[width=\textwidth]{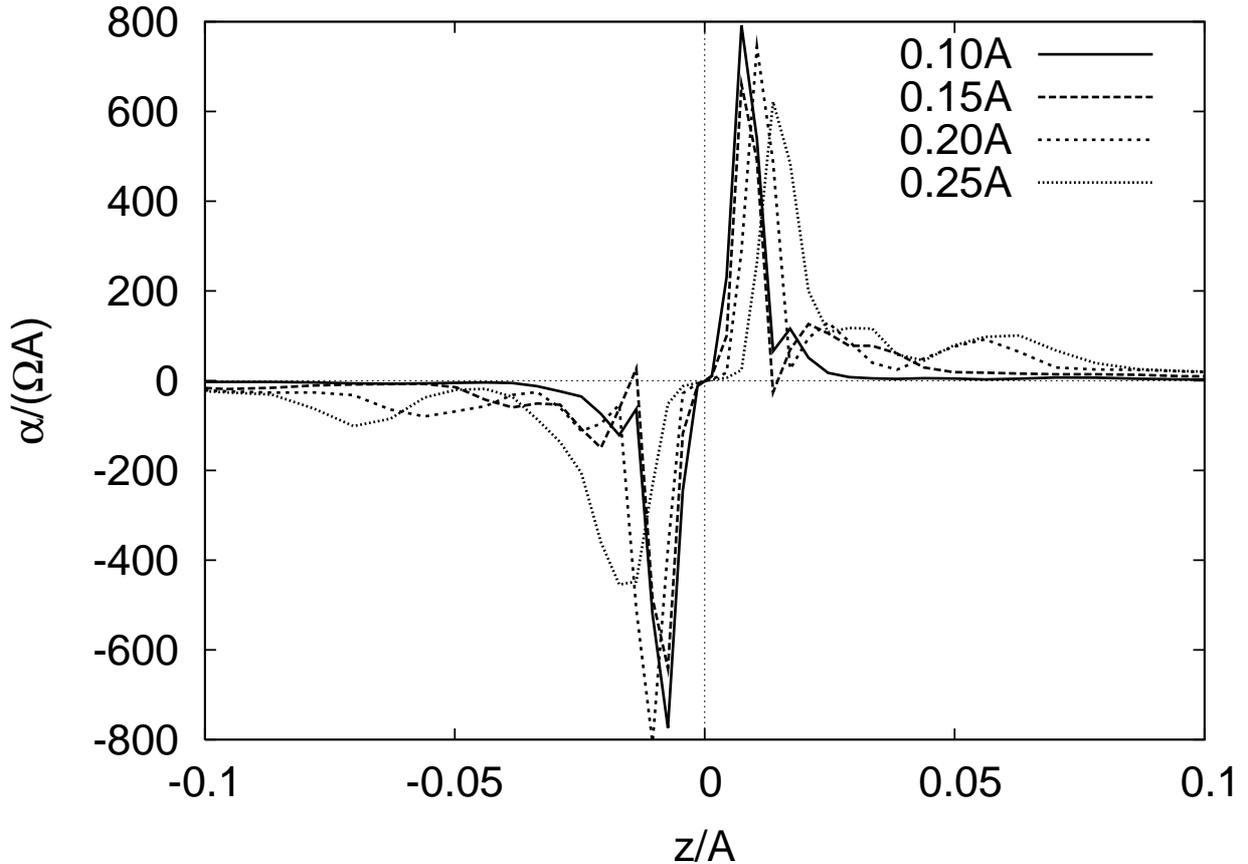}%
\caption{Distribution of the mean helicity $\alpha$ in the accretion disk at various distances from the accretor.}%
\label{fig_alpha}
\end{figure}

Figure \ref{fig_alpha} shows vertical distributions of $\alpha$ for non-azimuthal velocity fluctuations in the accretion disk at various distances from the accretor obtained via three-dimensional MHD numerical simulations. The helicity possesses a mirror antisymmetry about the plane of the disk, which is a necessary condition for the operation of the dynamo mechanism \cite{Ruzmaikin1988}. Similar profiles are probably obtained for turbulent velocity fluctuations as well. To simplify our further analysis, we will use values of $\alpha$ averaged over half the disk height. 

We will restrict our analysis to stationary solutions of (\ref{eq4.6}, \ref{eq4.7}). In this case, these equations can be reduced to a single equation for the toroidal field:
\begin{equation}\label{eq4.12}
  \pdiff{^3B_{\varphi}}{z^3} +
  \fr{\alpha g}{\eta^2} B_{\varphi} +
  \fr{g B_0}{\eta r} = 0.
\end{equation}
This equation is invariant to the transformation $z \to -z$ simultaneous with $B_{\varphi}(z) \to -B_{\varphi}(-z)$. This means that it describes the generation of dipolar (antisymmetric) magnetic field, for which $B_{\varphi}(-z) = -B_{\varphi}(z)$. As was shown in the previous section, this type of field is excited by the stellar field subject to the differential rotation of the disk. However, the quadrupolar (symmetric) field, for which $B_{\varphi}(-z) = B_{\varphi}(z)$, may become dominant at large distances from the star, where $\sfr{g B_0 H^3}/(\eta r) \ll 1$ \cite{Ruzmaikin1988}. 

Due to the linearity of (\ref{eq4.12}), a general solution can be written in the form $B_{\varphi} = B^{(d)}_{\varphi} + B^{(q)}_{\varphi}$, where $B^{(d)}_{\varphi}$ and $B^{(q)}_{\varphi}$ describe the dipolar and quadrupolar field components and satisfy the equations
\begin{equation}\label{eq4.13}
  \pdiff{^3B^{(d)}_{\varphi}}{z^3} +
  \fr{\alpha g}{\eta^2} B^{(d)}_{\varphi} +
  \fr{g B_0}{\eta r} = 0,
\end{equation}
\begin{equation}\label{eq4.14}
  \pdiff{^3B^{(q)}_{\varphi}}{z^3} +
  \fr{\alpha g}{\eta^2} B^{(q)}_{\varphi} = 0.
\end{equation}

These equations can be rewritten in dimensionless form. We denote $B^{(d,q)}_{\varphi} = B_{\varphi,0} f_{d,q}(\zeta)$, where $\zeta = \sfr{z}{H}$. Equations (\ref{eq4.13}, \ref{eq4.14}) then take the form
\begin{equation}\label{eq4.15}
  {f_d}''' + D f_d + 1 = 0,
\end{equation}
\begin{equation}\label{eq4.16}
  {f_q}''' + D f_q = 0,
\end{equation}
where $D = \sfr{\alpha g H^3}{\eta^2}$ is the dynamo number and a prime denotes differentiation with respect to $\zeta$. These equations can be solved using the boundary conditions $f_d(0) = f_d(1) = 0$, ${f_d}''(1) = -1$, ${f_q}'(0) = f_q(1) = {f_q}''(1) = 0$. The dynamo number $D$ should be negative in outer parts of the disk when $z > 0$ since the differential-rotation measure $g$ is negative in these regions. 

The solution of (\ref{eq4.15}) for the dipolar mode can be written
\begin{equation}\label{eq4.17}
  f_d = \fr{1}{N^3} + C_d^{(1)} e^{N\zeta} + 
  e^{-\frac{N}{2}\zeta}
  \left[
   C_d^{(2)} \sin\left( \fr{\sqrt{3}N}{2}\zeta \right) + 
   C_d^{(3)} \cos\left( \fr{\sqrt{3}N}{2}\zeta \right)
  \right],
\end{equation}
where $N = |D|^{\sfr{1}{3}}$. The constants of integration $C_d^{1,2,3}$ in this expression can be found from the boundary conditions. In the limit of small dynamo numbers as $N \to 0$, we obtain the asymptotic relation $f_d = \zeta(1 - \zeta^2)/6 + O(N)$, corresponding to the solution (\ref{eq4.10}).

The solution of (\ref{eq4.16}) for the quadrupolar mode can be written \cite{Campbell1997}
\begin{equation}\label{eq4.18}
  f_q = C_q 
  \left[
   e^{N(\zeta - 1)} - 
   2 e^{-\frac{N}{2}(\zeta - 1)}
   \cos\left( \fr{\sqrt{3}N}{2}(\zeta - 1) + \fr{\pi}{3} \right)
  \right],
\end{equation}
where $C_q$ is the amplitude. The quantity $N$ satisfies the relation
\begin{equation}\label{eq4.19}
  2 \cos\left( \fr{\sqrt{3}N}{2} \right) + 
  e^{-\frac{3N}{2}} = 0.
\end{equation}
The roots of this equation determine the discrete set of allowed values of $N$. The first three roots are $N_1 = 1.85$, $N_2 = 5.44$, $N_3 = 9.07$.

The obtained dependences (\ref{eq4.17}) and (\ref{eq4.18}) determine the vertical structure of the field. Note that, when the dynamo effect is included, the radial structure of the field is described by $B_{\varphi,0}$, as before. The quadrupolar component will probably always dominate in the inner magnetic ring when the magnetic axis is inclined. The quadrupolar field can also become dominant in the outer magnetic ring with time, since this component grows more rapidly than the dipolar
field due to the dynamo effect \cite{Ruzmaikin1988}. This magnetic-field structure in the inner and outer rings is clearly
demonstrated by our computations. 

\section{Conclusion}

We have investigated the magnetic-field structure in accretion disks in semi-detached binary systems based on three-dimensional numerical simulations. These assumed that the intrinsic magnetic field of the accretor is dipolar, with the dipole axis, in general, inclined to the rotation axis. The numerical simulations took into account the diffusion of magnetic field due to the dissipation of currents in turbulent vortices and magnetic buoyancy. We considered the SS Cyg system as an example.

The results of the simulations show that the disk magnetic field is predominantly toroidal. Three zones can be distinguished in the disk: an inner zone with the intense generation of toroidal field via the differential rotation of the disk, a zone with current sheets, and an outer zone of dissipation of the magnetic field. A current sheet forms in the inner zone, due to the variation of the rotation law for the disk material near the stellar magnetosphere. Further, this current sheet is carried into the outer part of the disk with decreted material, with a new current sheet forming in its place after some time. Therefore, several current sheets can be present simultaneously in the intermediate zone. The alternation of accretion and decretion regimes in the inner parts of the disk is probably due to the generation of magnetic field and the associated increase in the corresponding magnetic-pressure gradient, which acts to stop the accretion. This is followed by an outflow of the field into the outer part of the disk, decreasing the magnetic pressure and bringing about a transition back to an accretion regime.

\begin{figure}[t]
\centering
\includegraphics[width=\textwidth]{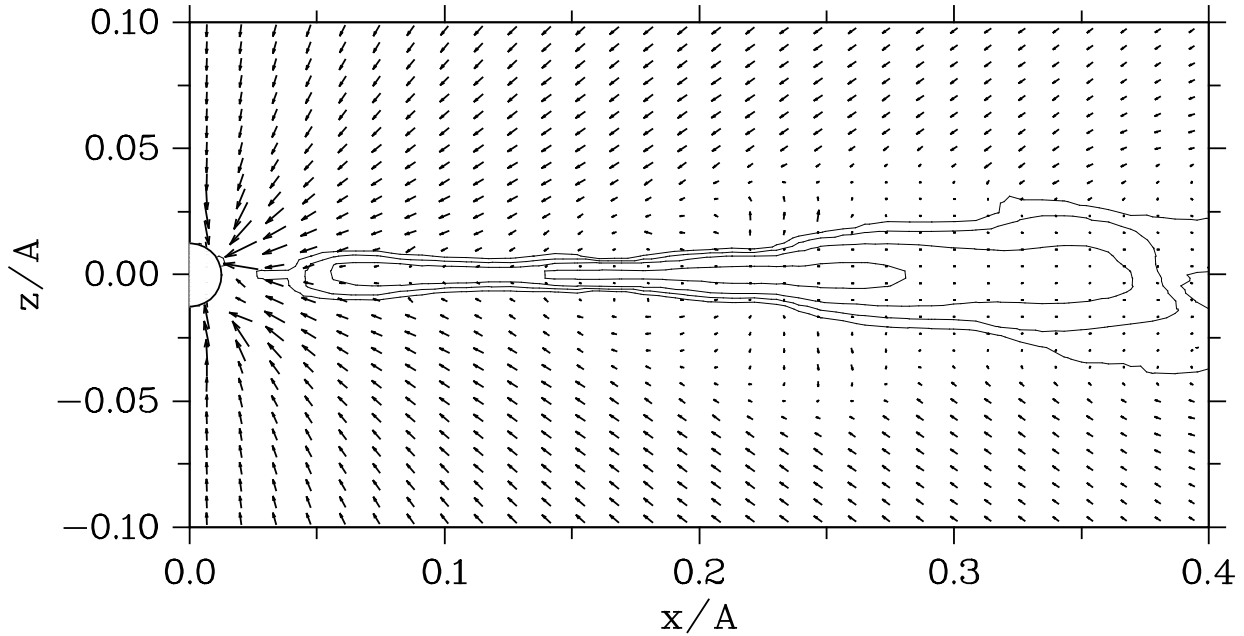}%
\caption{Distribution of the density and velocity in the $xz$ plane.}%
\label{fig_rhoxz}
\end{figure}

Current sheets should form in systems in which the corotation radius exceeds the magnetosphere radius. This condition is determined by the relationship between the accretor’s rotational angular velocity, magnetic field, and accretion rate. The field dissipation and temperature are enhanced in current sheets, leading to the emission of UV and X-ray radiation. Moreover, the total pressure is reduced in current sheets due to the decrease in the magnetic field. Therefore, the disk thickness should decrease in the vicinity of current sheets, as is confirmed by the numerical simulations. Figure  \ref{fig_rhoxz} shows the density distribution in the $xz$ plane in the disk at time $13.36 P_{\text{orb}}$. The disk thickness decreases slightly in the intermediate zone ($0.1 \le r/A \le 0.2$), where the current sheets are located (Fig. \ref{fig_lgb}), compared to the inner and outer zones. Note that, in the purely gas-dynamic case \cite{Zhilkin2009} the disk
thickness increases monotonically with distance. 

The magnetic field in the inner zone interacts intensely with a spiral precessional wave. As a result, the accretion rate grows when the precessional wave approaches the stellar surface near the magnetic poles. This gives rise to a flare in the accretion rate (with an amplitude of the order of $15\%$) twice per period, when the spiral wave passes the North and South magnetic poles of the star. This behavior of the accretion rate is in good agreement with the observed UV flux variations of SS Cyg between outbursts.

A vertical gradient of the field arises in the disk when the magnetic axis is inclined. Therefore, the generation of toroidal field in the disk acquires a symmetrical (quadrupolar) character about the equatorial plane. This effect is most strongly expressed when the inclination of the magnetic axis is $\theta=45^{\circ}$. An antisymmetric toroidal field (dipolar mode) will be generated only when $\theta=0^{\circ}$ and $90^{\circ}$. Small deviations from these values lead to a suppression of the dipolar mode by the quadrupolar mode. Thus, the generation of a purely dipolar mode in the accretion disks of intermediate polars may be unstable. The quadrupolar mode is more stable.

Our computations show that motion in the accretion disk displays a mirror antisymmetrical helicity. Therefore, the magnetic field in the outer parts of the disk may be partially generated by the dynamo mechanism. Here, the quadrupolar mode again grows most rapidly. However, the role of the dynamo is small, since dynamical effects dominate in the disk.\\

This work was supported by the Basic Research Program of the Presidium of the Russian Academy of Sciences on ''The Origin, Structure, and Evolution of Objects in the Universe'', the Russian Foundation for Basic Research (projects 08-02-00371, 09-02-00064), the Federal Targeted Program ''Science and Science Education for Innovation in Russia 2009–2013'' (a grant of the Federal Agency on Science and Innovation of the Ministry of Science and Education of the Russian Federation ''Studies of Non-Stationary Processes in Stars and the Interstellar Medium in the Institute of Astronomy of the Russian Academy of Sciences'').

\end{document}